\documentclass{ws-p8-50x6-00}
\usepackage{feynmp}                       %%% for Feynman diagrams; a few
\usepackage[dvips]{color}                 %%% for colours in the graphs
\usepackage{graphicx}                     %%% for pictures and figures
\usepackage{amssymb}

\newcommand{\absatz}{}

\newcommand{\non}{\nonumber}

\newcommand{\ii}{{\rm i}}

\newcommand{\tr}{{\rm tr}}
\newcommand{\T}{{\rm T}}

\newcommand{\mpi}{m_\pi}
\newcommand{\fpi}{f_\pi}
\newcommand{\MeV}{{\rm MeV}}
\newcommand{\de}{\partial}
\newcommand{\dev}{\vec{\de}}

\begin{document}

\begin{fmffile}{bolfeyn}
  \fmfset{curly_len}{2mm} \fmfset{dash_len}{1.5mm} \fmfset{wiggly_len}{3mm}
  \newcommand{\feynbox}[2]{\mbox{\parbox{#1}{#2}}}
  \newcommand{\fs}{\scriptstyle}
                          %%%%%% adjusts the size of labels in feynmf-diagrams
  \newcommand{\hq}{\hspace{0.5em}} \newcommand{\hqm}{\hspace{-0.25em}}
  
  \fmfcmd{vardef ellipseraw (expr p, ang) = save radx; numeric radx; radx=6/10
    length p; save rady; numeric rady; rady=3/10 length p; pair center;
    center:=point 1/2 length(p) of p; save t; transform t; t:=identity xscaled
    (2*radx*h) yscaled (2*rady*h) rotated (ang + angle direction length(p)/2
    of p) shifted center; fullcircle transformed t enddef;
    style_def ellipse expr p= shadedraw ellipseraw (p,0); enddef; }

%%%%%%%%%%%%%%%%%%%%%%%%%%%%%%%%%%%%%%%%%%%%%%%%%%%%%%%%%%%%%%%%%%%%%%%%%%%%%%%
%%%%%%%%%%%%%%%%%%%%%%%%%%%%%%%%%%%%%%%%%%%%%%%%%%%%%%%%%%%%%%%%%%%%%%%%%%%%%%%
    
  \title{A Tale of Two and Three Bodies\thanks{Talk held at the Conference
      Bologna 2000 - Structure of the Nucleus at the Dawn of the Century,
      Bologna (Italy) 29th May -- 3rd June 2000; to be published in the
      Proceedings; preprint numbers nucl-th/0009059, TUM-T39-00-16.}  }
  
  \author{Harald W.~Grie\3hammer}
  
  \address{
    Institut f{\"u}r Theoretische Physik, Physik-Department der\\
    Technischen Universit{\"a}t M{\"u}nchen, 85748 Garching, Germany\\
    Email: hgrie@physik.tu-muenchen.de}

  \maketitle

%%%%%%%%%%%%%%%%%%%%%%%%%%%%%%%%%%%%%%%%%%%%%%%%%%%%%%%%%%%%%%%%%%%%%%%%%%%%%%%
%%%%%%%%%%%%%%%%%%%%%%%%%%%%%%%%%%%%%%%%%%%%%%%%%%%%%%%%%%%%%%%%%%%%%%%%%%%%%%%
%%%%%%%%%%%%%%%%%%%%%%%%%%%%%%%%%%%%%%%%%%%%%%%%%%%%%%%%%%%%%%%%%%%%%%%%%%%%%%%
% Main Body
%
\noindent
This presentation is a concise teaser for the Effective Field Theory (EFT) of
two and three nucleon systems as it emerged in the last three years, using a
lot of words and figures, and a few cheats. For details, I refer to the
exhaustive bibliographies in\cite{INTWorkshopSummary}, and papers with
J.-W.~Chen, R.P.~Springer and M.J.~Savage\cite{Compton},
P.F.~Bedaque\cite{pbhg,pbfghg}, and F.~Gabbiani\cite{pbfghg}.

%\absatz
Effective Field Theory methods are largely used in many branches of physics
where a separation of scales exists. In low energy nuclear systems, the scales
are, on one side, the low scales of the typical momentum of the process
considered and the pion mass, and on the other side the higher scales
associated with chiral symmetry and confinement. This separation of scales
produces a low energy expansion, resulting in a description of strongly
interacting particles which is systematic, rigorous and model independent
(meaning, independent of assumptions about the non-perturbative QCD dynamics).

\absatz Three main ingredients enter the construction of an EFT: The
Lagrangean, the power counting and a regularisation scheme.  First, the
relevant degrees of freedom have to be identified. In his original suggestion
how to extend EFT methods to systems containing two or more nucleons,
Weinberg\cite{Weinberg} noticed that below the $\Delta$ production scale, only
nucleons and pions need to be retained as the infrared relevant degrees of
freedom of low energy QCD.  The theory becomes non-relativistic at leading
order in the velocity expansion, with relativistic corrections systematically
included at higher orders. The most general chirally (and iso-spin) invariant
Lagrangean consists hence of contact interactions between non-relativistic
nucleons, and between nucleons and pions, with the first few terms of the form
\begin{eqnarray}\label{ksw}
   {\mathcal{L}}_{NN}&=&N^{\dagger}(\ii \de_0+\frac{\dev^2}{2M})N+
   \;\frac{\fpi^2}{8}\;
   \tr[(\partial_{\mu} \Sigma^{\dagger})( \partial^{\mu} \Sigma)]\;+
   \;g_A N^{\dagger} \vec{A}\cdot\sigma N\;-\non\\
   &&-\;C_0 (N^{\T} P^i N)^{\dagger} \ (N^{\T} P^i N)\;
   +\\&&
   + \;\frac{C_2}{8}
   \left[(N^{\T} P^i N)^{\dagger} (N^{\T} P^i
     (\stackrel{\scriptscriptstyle\rightarrow}{\de}-
      \stackrel{\scriptscriptstyle\leftarrow}{\de})^2 N)+
   {\rm H.c.}\right]
   + \dots\;\;,\nonumber
\end{eqnarray}
where $N={p\choose n}$ is the nucleon doublet of two-component spinors and
$P^i$ is the projector onto the iso-scalar-vector channel, $
P^{i,\,b\beta}_{a\alpha}=\frac{1}{\sqrt{8}}
(\sigma_2\sigma^i)_{\alpha}^{\beta} (\tau_2)_a^b$.  The iso-vector-scalar part
of the $NN$ Lagrangean introduces more constants $C_i$ and interactions and
has not been displayed for convenience. The field $\xi$ describes the pion,
$\xi(x)=\sqrt{\Sigma}=e^{\ii \Pi/\fpi}$, $
A_{\mu}=\frac{\ii}{2}(\xi\partial_{\mu}\xi^{\dagger}-\xi^{\dagger}
\partial_{\mu}\xi)$. All short distance physics -- branes and strings, quarks
and gluons, resonances like the $\Delta$ or $\sigma$ -- is integrated out into
the coefficients of the low energy Lagrangean. The most practical way to
determine those constants is by fitting them to experiment.  The EFT with
pions integrated out (formally, $g_A=0$ in (\ref{ksw})) is valid below the
pion cut and was recently pushed to very high orders in the two-nucleon
sector\cite{CRS} where accuracies of the order of $1\%$ were obtained. It can
be viewed as a systematisation of Effective Range Theory with the inclusion of
relativistic and short distance effects traditionally left out in that
approach.

\absatz As the second part of an EFT formulation, predictive power is ensured
by establishing a power counting scheme, i.e.\ a way to determine at which
order in a momentum expansion different contributions will appear, and keeping
only and all the terms up to a given order. The dimensionless, small parameter
on which the expansion is based is the typical momentum $Q$ of the process in
units of the scale $\Lambda$ at which the theory is expected to break down.
Values for $\Lambda$ and $Q$ have to be determined from comparison to
experiments and are a priori unknown. Assuming that all contributions are of
natural size, i.e.\ ordered by powers of $Q$, the systematic power counting
ensures that the sum of all terms left out when calculating to a certain order
in $Q$ is smaller than the last order retained, allowing for an error estimate
of the final result.

Even if calculations of nuclear properties were possible starting from the
underlying QCD Lagrangean, EFT simplifies the problem considerably by
factorising it into a short distance part (subsumed into the coefficient of
the Lagrangean) and a long distance part which contains the infrared-relevant
physics and is dealt with by EFT methods. EFT provides an answer of finite
accuracy because higher order corrections are systematically calculable and
suppressed in powers of $Q$. Hence, the power counting allows for an error
estimate of the final result. Relativistic effects, chiral dynamics and
external currents are included systematically, and extensions to include
e.g.~parity violating effects are straightforward. Gauged interactions and
exchange currents are unambiguous.  Results obtained with EFT are easily
dissected for the relative importance of the various terms.  Because only
$S$-matrix elements between on-shell states are observables, ambiguities
nesting in ``off-shell effects'' are absent.  On the other hand, because only
symmetry considerations enter the construction of the Lagrangean, EFTs are
less restrictive as no assumption about the underlying QCD dynamics is
incorporated.

In systems involving two or more nucleons, establishing a power counting is
complicated by the fact that unnaturally large scales have to be accommodated:
Given that the typical low energy scale in the problem should be the mass of
the pion as the lightest particle emerging from QCD, fine tuning seems to be
required to produce the large scattering lengths in the ${}^1{\rm S}_0$ and
${}^3{\rm S}_1$ channels ($1/a^{{}^1{\rm S}_0}=-8.3\;\MeV,\; 1/a^{{}^3{\rm
    S}_1}=36\;\MeV$). Since there is a bound state in the ${}^3{\rm S}_1$
channel with a binding energy $B=2.225\;\MeV$ and hence a typical binding
momentum $\gamma=\sqrt{M B}\simeq 46\;\MeV$ well below the scale $\Lambda$ at
which the theory should break down, it is also clear that at least some
processes have to be treated non-perturbatively in order to accommodate the
deuteron.  A way to incorporate this fine tuning into the power counting was
suggested by Kaplan, Savage and Wise\cite{KSW}: At very low momenta, contact
interactions with several derivatives -- like $p^2C_2$ and the pion-nucleon
interactions -- should become unimportant, and we are left only with the
contact interactions proportional to $C_0$. The leading order contribution to
nucleons scattering in an ${\rm S}$ wave comes hence from four nucleon contact
interactions and is summed geometrically, and the coefficient of the
four-nucleon interactions scale as $C_0\sim\frac{1}{M Q}$, $C_2\sim\frac{1}{M
  \Lambda Q^2}$, \dots.  Dimensional regularisation preserves the systematic
power counting as well as all symmetries (esp.~chiral invariance) at each
order in every step of the calculation.  Even at NNLO in the two nucleon
system, simple, closed answers allow one to assert the analytic structure. The
deuteron propagator has the correct pole position and cut structure.

One surprising result arises from this analysis because chiral symmetry
implies a derivative coupling of the pion to the nucleon at leading order, so
that the instantaneous one pion exchange scales as $Q^0$ and is {\it smaller}
than the contact piece $C_0\sim Q^{-1}$.  Pion exchange and higher derivative
contact terms appear hence only as perturbations at higher orders.  In
contradistinction to iterative potential model approaches, each higher order
contribution is inserted only once.  In this scheme, the only non-perturbative
physics responsible for nuclear binding is extremely simple, and the more
complicated pion contributions are at each order given by a finite number of
diagrams.

\begin{figure}[!htb]
  \centerline{\epsfig{file=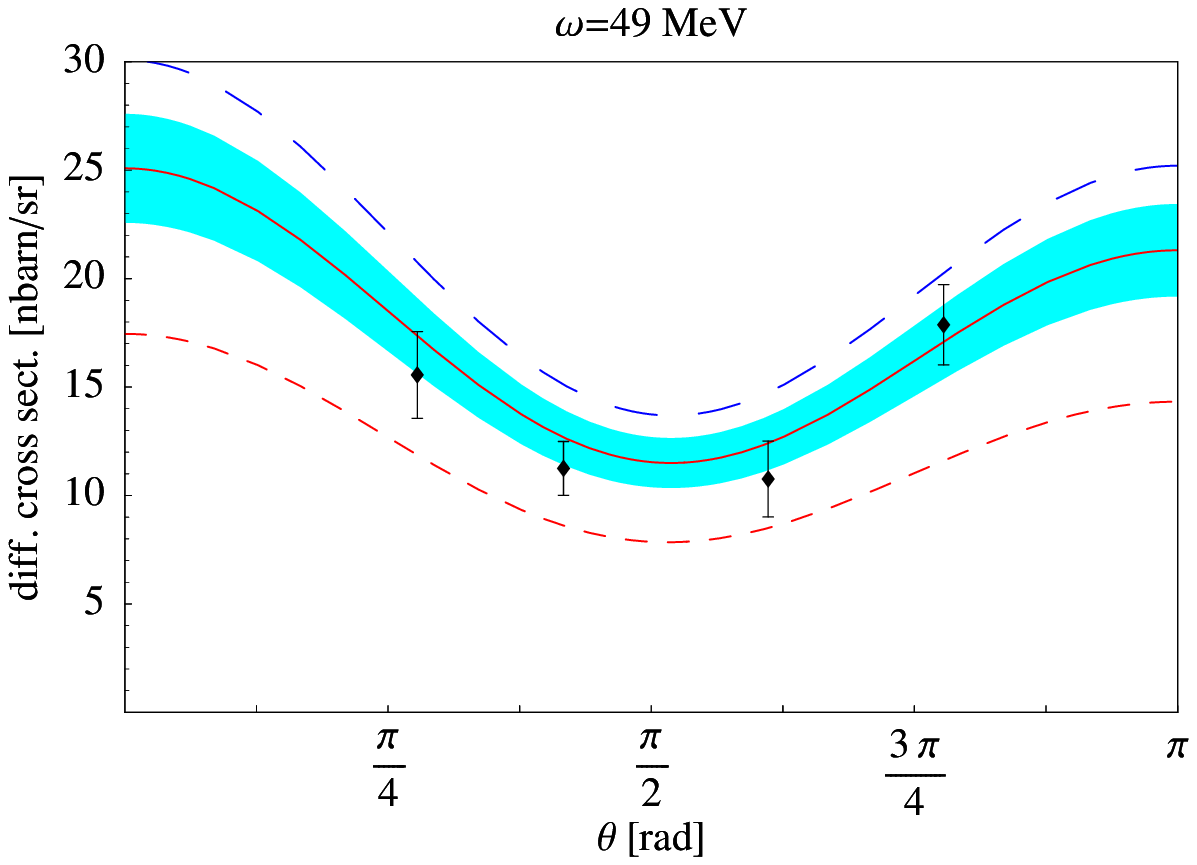,width=0.49\textwidth,clip=} \hfill
    \epsfig{file=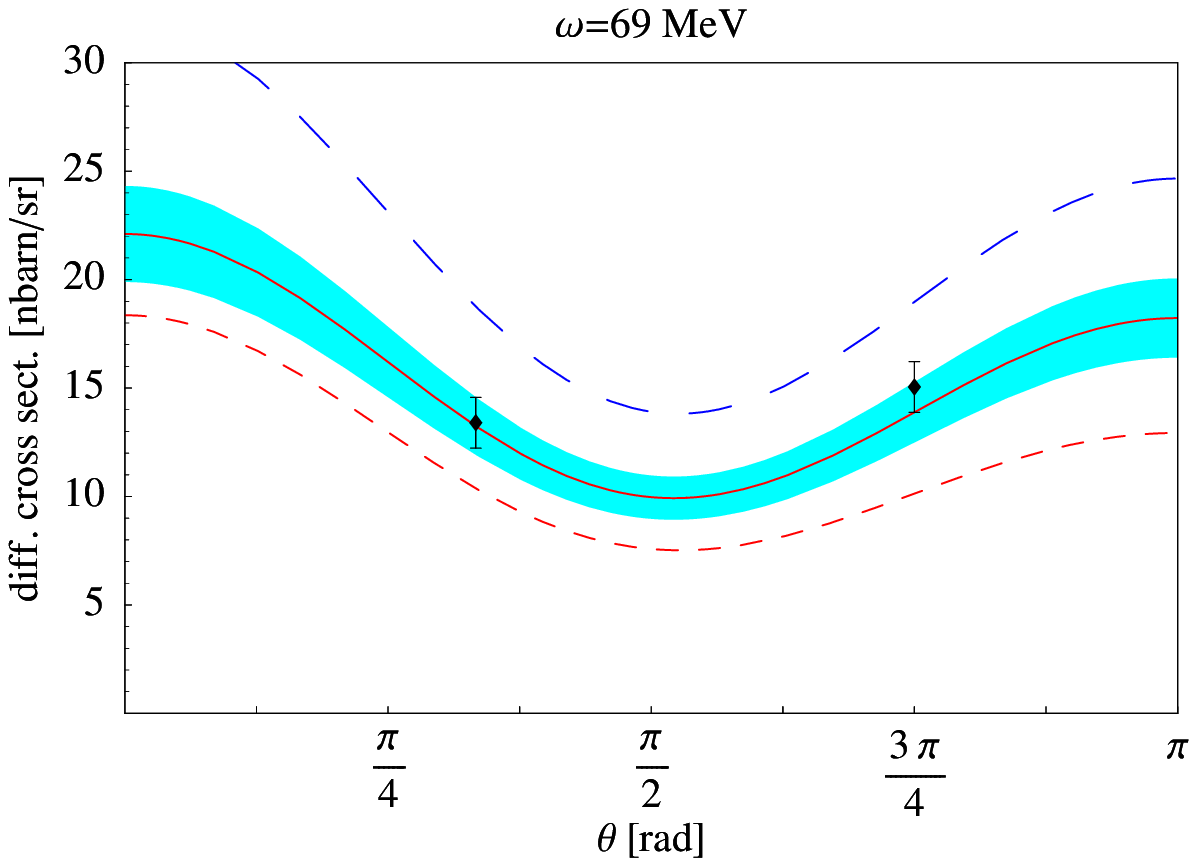,width=0.49\textwidth,clip=}}
\caption{The differential cross section for elastic 
  \protect$\gamma$-deuteron Compton scattering at incident photon energies of
  \protect$E_{\gamma}=49\ {\rm MeV}$ and \protect$69\ {\rm MeV}$ in an EFT
  with explicit pions\protect\cite{Compton}, no free parameters. Dashed: LO;
  long dahed: %dotted:
  NLO without the graphs that contribute to the nucleon polarisability; solid
  curve: complete NLO result. Accuracy of calculation at NLO (\protect$\pm
  10\%$) indicated by shaded area.}
\label{fig:compton}
\vspace*{-4ex}
\end{figure}

In this formulation, the elastic deuteron Compton scattering cross
section\cite{Compton} to NLO is parameter-free with an accuracy of $10\%$.
Contributions at NLO include the pion graphs that dominate the electric
polarisability of the nucleon from their $\frac{1}{m_\pi}$ behaviour in the
chiral limit. The comparison with experiment in Fig.~\ref{fig:compton} shows
good agreement and therefore confirms the HB$\chi$PT value for $\alpha_E$. The
deuteron scalar and tensor electric and magnetic polarisabilities are also
easily extracted\cite{Compton}.

%\noindent
In the three body sector, the equations that need to be solved are
computationally trivial, as opposed to many-dimensional integral equations
arising in other approaches.  The absence of Coulomb interactions in the $nd$
system ensures that only properties of the strong interactions are probed. In
the quartet channel, the Pauli principle forbids three body forces in the
first few orders.  In the ${\rm S}$ wave, spin-doublet (triton) channel, an
unusual renormalisation makes the three-body force large and as important as
the leading two-body forces\cite{Stooges2}. More work is needed there.

\begin{figure}[!htb]
  \begin{center}
    \setlength{\unitlength}{0.7pt}
    \feynbox{90\unitlength}{
            \begin{fmfgraph*}(90,50)
              \fmfleft{i2,i1} \fmfright{o2,o1}
              %\fmfv{label=$\fs(\frac{3\kv^2}{4M}-\frac{\gamma^2}{M},,\kv)$,
%                label.angle=90}{i1}
%              \fmfv{label=$\fs(\frac{3\kv^2}{4M}-\frac{\gamma^2}{M}+\epsilon
%                ,,\pv)$,label.angle=90}{o1}
%              \fmfv{label=$\fs(\frac{\kv^2}{2M},,-\kv)$,label.angle=-90}{i2}
%              \fmfv{label=$\fs(\frac{\kv^2}{2M}-\epsilon,,-\pv)$,
%                label.angle=-90}{o2}
              \fmf{double,width=thin,tension=3}{i1,v1}
              \fmf{double,width=thin,tension=1.5}{v1,v3,v2}
              \fmf{double,width=thin,tension=3}{v2,o1}
              \fmf{vanilla,width=thin}{i2,v4,o2} \fmffreeze \fmffreeze
              \fmf{ellipse,rubout=1}{v3,v4}
              \end{fmfgraph*}}
            \hq$=$\hq \feynbox{90\unitlength}{
            \begin{fmfgraph*}(90,50)
              \fmfleft{i2,i1} \fmfright{o2,o1}
              \fmf{double,width=thin,tension=4}{i1,v1,v2}
              \fmf{vanilla,width=thin}{v2,o1}
              \fmf{double,width=thin,tension=4}{v3,v4,o2}
              \fmf{vanilla,width=thin}{i2,v3} \fmffreeze
              \fmf{vanilla,width=thin}{v2,v3}
              \end{fmfgraph*}}
            \hq$+$\hq \feynbox{170\unitlength}{
            \begin{fmfgraph*}(170,50)
              \fmfleft{i2,i1} \fmfright{o2,o1}
              \fmf{double,width=thin,tension=3}{i1,v1}
              \fmf{double,width=thin,tension=1.5}{v1,v6,v5}
              \fmf{double,width=thin,tension=3}{v5,v2}
              \fmf{vanilla,width=thin}{v2,o1} \fmf{vanilla,width=thin}{i2,v7}
              \fmf{vanilla,width=thin,tension=0.666}{v7,v4}
              \fmf{double,width=thin,tension=4}{v4,v3,o2} \fmffreeze
              \fmf{vanilla,width=thin}{v4,v2} \fmf{ellipse,rubout=1}{v6,v7}
              \end{fmfgraph*}}
 %           \vspace{10pt}
    \caption{The Faddeev equation for the three body system.}
    \label{fig:LOfaddeev}
  \end{center}
  \vspace*{-3ex}
\end{figure}
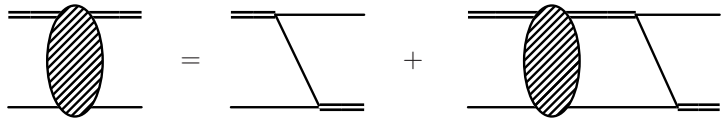

A comparative study between the theory with explicit pions and the one with
pions integrated out was performed\cite{pbhg} in the spin quartet ${\rm S}$
wave for momenta of up to $300\;\MeV$ in the centre-of-mass frame ($E_{{\rm
    cm}}\approx70\;\MeV$). As seen above, the two theories are identical at
LO: All graphs involving only $C_0$ interactions are of the same order and
form a double series which
%is not geometrical and
cannot be written down in closed form. Summing all ``bubble-chain'' sub-graphs
into the deuteron propagator, one can however obtain the solution numerically
from the integral equation pictorially shown in Fig~\ref{fig:LOfaddeev}.

\begin{figure}[!htb] 
  \centerline{\epsfig{file=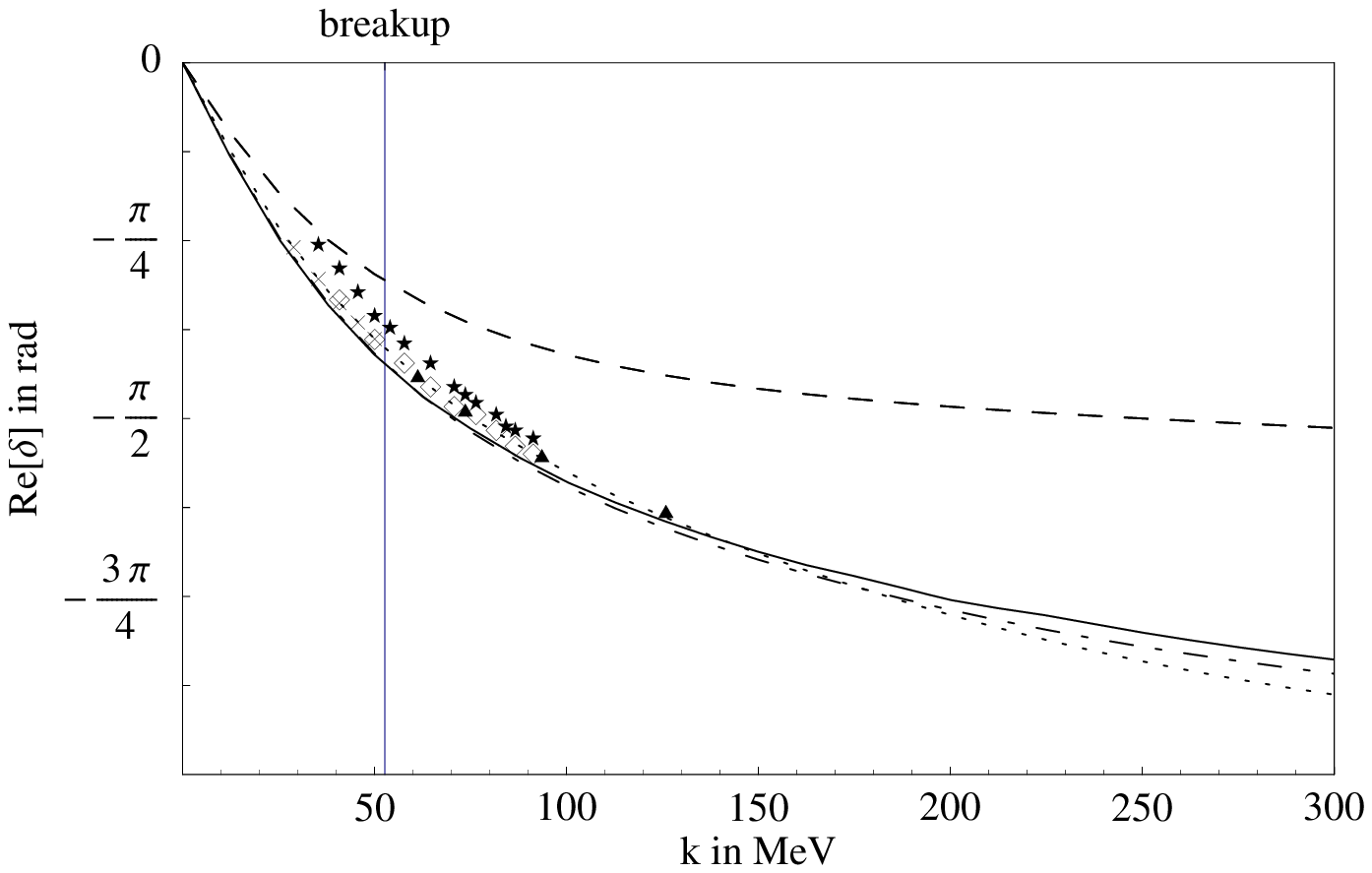,width=0.5\textwidth,clip=}
    \hspace{-1.5ex}
    \epsfig{file=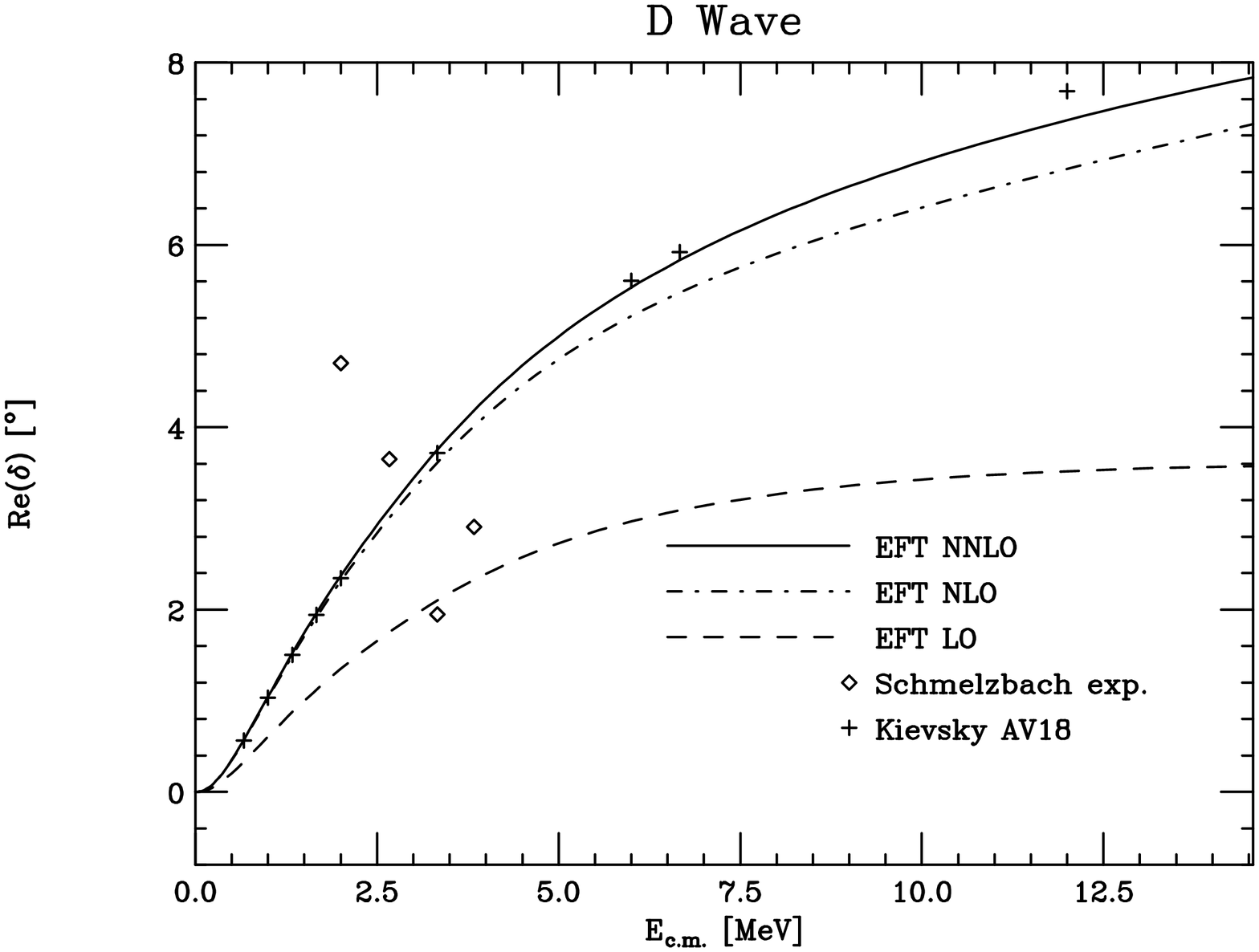,height=0.4\textwidth,angle=90,clip=}}
  %\vspace{10pt}
\caption{Real parts in the quartet \protect${\rm S}$ and doublet \protect${\rm
    D}$ wave phase shift of \protect$nd$ scattering versus the centre-of-mass
  momentum\protect\cite{pbhg,pbfghg}. Dashed: LO; solid (dot-dashed) line: NLO
  with perturbative pions (pions integrated out); dotted: NNLO without
  pions\protect\cite{pbfghg}. Realistic potential models: squares, crosses,
  triangles.  Stars: $pd$ phase shift analysis.}
\label{fig:delta}
\vspace*{-4ex}
\end{figure}

The calculation with/without explicit pions to NLO/NNLO shows convergence.
Pionic corrections to $nd$ scattering in the quartet ${\rm S}$ wave channel --
although formally NLO -- are indeed much weaker. The difference to the theory
in which pions are integrated out should appear for momenta of the order of
$\mpi$ and higher because of non-analytical contributions of the pion cut, but
those seem to be very moderate, see Fig.~\ref{fig:delta}.

%\noindent
Finally, the real and imaginary parts of the higher partial waves
$l=1,\dots,4$ in the spin quartet and doublet channel were found\cite{pbfghg}
in a blablameter-free calculation, see Fig.~\ref{fig:delta}.  Within the range
of validity of this pion-less theory, convergence is good, and the results
agree with potential model calculations (as available) within the theoretical
uncertainty. That makes one optimistic about carrying out higher order
calculations of problematic spin observables like the $A_y$ problem where the
EFT approach will differ from potential model calculations due to the
inclusion of three-body forces.

%%%%%%%%%%%%%%%%%%%%%%%%%%%%%%%%%%%%%%%%%%%%%%%%%%%%%%%%%%%%%%%%%%%%%%%%%%%%%%%
%%%%%%%%%%%%%%%%%%%%%%%%%%%%%%%%%%%%%%%%%%%%%%%%%%%%%%%%%%%%%%%%%%%%%%%%%%%%%%%
%%%%%%%%%%%%%%%%%%%%%%%%%%%%%%%%%%%%%%%%%%%%%%%%%%%%%%%%%%%%%%%%%%%%%%%%%%%%%%%

%%%%%%%%%%%%%%%%%%%%%%%%%%%%%%%%%%%%%%%%%%%%%%%%%%%%%%%%%%%%%%%%%%%%%%%%%%%%%%%
\end{fmffile}
\end{document}